\documentclass[12pt]{article}
\usepackage{graphicx}
\usepackage{enumerate}
\def\lsim{\mathrel{\vcenter{\hbox{$<$}\nointerlineskip\hbox{$\sim$}}}}
\newcommand{\be}{\begin{equation}}
\newcommand{\ee}{\end{equation}}
\newcommand{\beq}[1]{\begin{equation}\label{#1}}
\newcommand{\eeq}{\end{equation}}
\newcommand{\bea}[1]{\begin{eqnarray} \label{#1}}
\newcommand{\eea}{\end{eqnarray}}
\newcommand{\ba}{\begin{array}}
\newcommand{\ea}{\end{array}}

\newcommand{\rf}[1]{(\ref{#1})}

\def\21{$SU(2) \otimes U(1) $}

\def\lsim{\raise0.3ex\hbox{$\;<$\kern-0.75em\raise-1.1ex\hbox{$\sim\;$}}}
\def\gsim{\raise0.3ex\hbox{$\;>$\kern-0.75em\raise-1.1ex\hbox{$\sim\;$}}} 

\newcommand{\mx}{\left[\begin{array}}
\newcommand{\finmx}{\end{array}\right]} 
\newcommand{\mxp}{\left(\begin{array}} 
\newcommand{\finmxp}{\end{array}\right)} 
\def\beq{\begin{equation}}
\def\eeq{\end{equation}}
\def\bea{\begin{eqnarray}}
\def\eea{\end{eqnarray}}

\def\mathbf#1{\hbox{\bf #1}}
\def\textrm#1{\hbox{#1}}
\def\half{{\textstyle{1 \over 2}}} 
 
\def\lsim{\raise0.3ex\hbox{$\;<$\kern-0.75em\raise-1.1ex\hbox{$\sim\;$}}}
\def\gsim{\raise0.3ex\hbox{$\;>$\kern-0.75em\raise-1.1ex\hbox{$\sim\;$}}}
\def\half{\frac{1}{2}}

\def\dmsq{\delta m^2}

\newcommand{\twidtheta}{\tilde{\theta}}
\newcommand{\eps}{\epsilon}
\newcommand{\Eres}{E_{\rm res}}

\begin{document}
\vspace*{-1in}
\renewcommand{\thefootnote}{\fnsymbol{footnote}}
\begin{flushright}
\texttt{
} 
\end{flushright}
\vskip 5pt
\begin{center}
{\Large{\bf Sterile-active neutrino oscillations 
and shortcuts in the extra dimension}
}
\vskip 25pt 

{\sf
Heinrich P\"as$^{1}$, Sandip Pakvasa$^{1}$, Thomas J. Weiler$^{2}$}
\vskip 10pt
{\it \small  
$^1$~Department of Physics \& Astronomy, 
University of Hawaii at Manoa,
2505 Correa Road, Honolulu, HI 96822, USA}\\
{\it \small  
$^2$~
Department of Physics and Astronomy, 
Vanderbilt University, Nashville, TN 37235, USA}\\

\vskip 20pt

{\bf Abstract}
\end{center}

\begin{quotation}
{\small
We discuss a possible new resonance in active-sterile neutrino oscillations
arising in theories with large extra dimensions.
Fluctuations in the brane effectively increase the path-length of 
active neutrinos relative to the path-length of sterile neutrinos 
through the extra-dimensional bulk. 
Well below the resonance, the standard oscillation formulas apply.
Well above the resonance, active-sterile oscillations are suppressed.
We show that a resonance energy in the range of $30-400$~MeV
allows an explanation of all neutrino oscillation data,
including LSND data, in a consistent four-neutrino model. 
A high resonance energy implies an enhanced signal in MiniBooNE\@.
A low resonance energy implies a distorted energy spectrum in LSND,
and an enhanced $\nu_\mu$~depletion from a stopped-pion source.
The numerical value of the resonance energy may be related back
to the geometric aspects of the brane world.
Some astrophysical and cosmological consequences of the brane-bulk 
resonance are briefly sketched.
}
\end{quotation}

\vskip 20pt  

\setcounter{footnote}{0}
\renewcommand{\thefootnote}{\arabic{footnote}}


\section{Introduction}
\label{sec:intro}

Theories with large extra dimensions  
typically confine the Standard Model (SM) particles on 
a 3+1 brane embedded in an extra-dimensional bulk \cite{xtra,RS}
see also \cite{Cremades:2002dh}.  
Gauge singlet particles may travel
on or off the brane.  These include the graviton, 
and any singlet ($\equiv$~``sterile'') neutrinos \cite{dvali}.
Virtual gravitons, too, penetrate the bulk, and so lead via Gauss' Law to
an apparent weak gravity on our brane, when in fact the strength of gravity 
may unify with the SM forces at scales as low as a few TeV.
Here we focus on the ``other'' possible particle in the bulk, the 
sterile neutrino \cite{sterilerev}.

The only evidence to date for the existence of the sterile neutrino comes
from the incompatibility of all reported 
neutrino oscillation results with the three-active neutrino world.
The solar and atmospheric disappearance data are corroborated,
whereas the LSND appearance data \cite{lsnd} are not yet corroborated. 
We will assume that the LSND data is correct,
and use this as motivation to study the possible compatibility of all the data
when the higher dimensional bulk is included.
We find a new active-sterile resonance which relates 
bulk and brane travel times.
When the new resonance energy falls between the LSND energies and the 
CDHS energies, 
then all the oscillation data become compatible.

This article is organized as follows: In section 2 we discuss a metric
with small-scale fluctuations, which allows for bulk shortcuts. 
In section 3 the effect of bulk 
shortcuts on active-sterile neutrino oscillations is illustrated in a 
simple model with one sterile and one active neutrino. Section 4 discusses
the LSND result in the context of constraints from other neutrino experiments
in a realistic 3+1 neutrino model. Some further
astrophysical and phenomenological 
constraints are discussed in section 5, and conclusions are drawn in section 
6.

\section{A metric for bulk shortcuts
\label{warpmetric}}

It has been shown that
branes embedded in higher dimensional spacetime are curved
extrinsically by self-gravity effects 
in the presence of matter
\cite{ishihara}. 
For example, while the on-brane distance between atomic constituents is 
fixed by electromagnetism 
and the Pauli principle, the attractive force of gravity between 
constituents can shorten the embedding distance of the brane in the bulk, 
leading naturally to
a scenario where the brane is deformed
(possesses fluctuations or ``buckles'') on
a microscopic scale. 
Alternative causes of brane bending include thermal and quantum fluctuations.
This picture leads to a framework
in which the on-brane geodesic felt by an active neutrino is longer
than the bulk geodesic felt by a sterile neutrino. 
Such apparent superluminal behavior for gauge-singlet quanta 
has been noted before, for the graviton~\cite{kaelb,freese}.

\begin{figure}
\centering
\includegraphics[clip,scale=0.50]{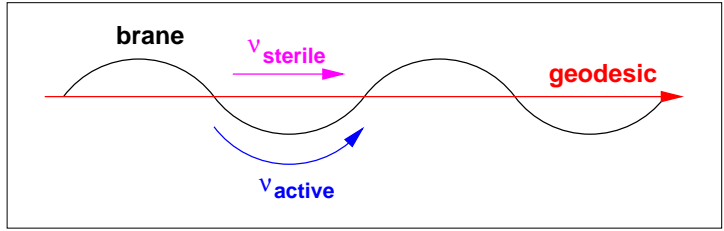}
\caption{Schematic representation of a periodically curved brane in 
Minkowski spacetime. A coordinate transformation leads to an 
equivalent description as a non-diagonal metric
with a flat brane, as described in the text.
}
 \label{fig:branebulk}
\end{figure}

In the spirit of \cite{freese} it is straightforward to construct a 
1+1 dimensional toy
model with a metric
which indeed exhibits the anticipated behavior.
First, write down a 1+2 dimensional embedding spacetime, 
with Minkowski metric
\be{}\label{flatmetric}
ds^2=dt^2-dx^2-dy^2\,.
\ee
In this embedding spacetime, assume 
the brane exhibits periodic (for simplicity) oscillations in space 
\be{}
y=A\,\sin k x\,;
\ee
see Fig.~\ref{fig:branebulk} for a schematic representation.
Here $k$ is the wave number of the fluctuation (in the $x$-direction),
and $A$ is the amplitude of the fluctuation (in the $y$-direction).
A brane with tension is dynamical,
and fluctuations change in time.
The toy-model fluctuations should be thought of as some rms average over
many time-slices.

The bulk-geodesic for the sterile neutrino is simply given by
$y=0$, which leads to a travel distance of
\be{}
D_g=x\,.
\ee
The geodesic for the active state on the brane is slightly more complicated:
\be{}
D_b = \int_{\rm brane} \sqrt{ dx^2 +dy^2} = \int^x \sqrt{1+A^2 k^2 \cos^2 k x}~dx\,.
\ee
We use subscripts $b$ and $g$ to denote the brane and bulk spaces, 
respectively.

In terms of the coordinate $x$, the parameter describing the shortcut in 
the bulk is 
\be{}\label{eps1}
\eps (x) = \frac{D_b -D_g}{D_b} = 
	1-\frac{x}{\int^x \sqrt{1+A^2 k^2 \cos^2 k x}~dx}\,.
\ee
While mathematically correct, this description of the geodesics as functions
of $x$ has a shortcoming, in that $x$ is not a coordinate easily identified 
in an experiment on the brane.
It is useful to consider a more physical set of brane coordinates.
They will lead to essentially the same parameter $\eps$.

Consider the space-coordinate transformation
\be{}
u=y-A\,\sin k x
\label{utrans}
\ee
and
\be{}
z= \int^x \sqrt{1+A^2 k^2 \cos^2 k x}~dx\,.
\label{ztrans}
\ee
Under this transformation, the line element in (\ref{flatmetric}) 
transforms into
\be{}
ds^2 = dt^2 - dz^2 - du^2 
  - \frac{2\,A\,k\, \cos k x(z)}{\sqrt{1+A^2 k^2 \cos^2 k x(z)}}~du~dz.
\label{metric}
\ee
Note that in $(u,z)$ coordinates, $u=0$ defines the location of the brane,
and $z$ labels the physical distance along the brane.
Consequently,  a photon moving along the brane ($u=0$) satisfies the equation
\be{}
ds^2=dt^2-dz^2\,,
\ee
and thus travels in time $t_f$ the distance
\be{}\label{branedistance}
z_b=t_f\,.
\ee

On the other hand, in $(x,y)$ coordinates, the brane is described by the 
periodic sine function, while the geodesic in the bulk follows a 
straight line
along mean $y$, given by
\be{}
y_g=0,~~~D_g=t.
\label{geoeq} 
\ee
Using eqs.\ (\ref{utrans}) and (\ref{ztrans}), the bulk geodesic 
equation (\ref{geoeq})
can be transformed into the
$(u,z)$ system,
\be{}
u_g=-A\,\sin(k t),
\label{ug}
\ee
\be{}
z_g=\int_0^{t_f} \sqrt{1+A^2 k^2 \cos^2 k t}~dt = 
\frac{\sqrt{1+A^2 k^2}}{k}
\,\,{\cal E}\left(k t_f,\sqrt{\frac{A^2 k^2}{1+ A^2 k^2}}\,\,\right),
\label{integral}
\ee
where ${\cal E}(p,q)$ denotes the elliptic integral of the second kind.

The bulk geodesic intersects the brane at $u_g =0$,
which according to eq.~(\ref{ug}) occurs at the discrete times 
\be{}
t_{\rm int} = \frac{n\pi}{k},
\label{tint}
\ee
where $n$ is an integer.
However, if the size of the brane's fluctuations is small on the scale of 
an experimental detector,
then it is not required that the two geodesics be in intersection,
and $t_{\rm int}$ has no special significance.

From a comparison of the integrand in (\ref{integral})
to the result of (\ref{branedistance}), one readily infers 
that $z_g>z_b$, which means that in a common time interval the bulk test 
particle seemingly travels farther in the 
physical $z$-coordinate than the brane particle.
In other words, the specific metric
(\ref{metric}) allows apparent superluminal propagation.
The shortcut in the bulk can be parametrized by 
\be{}\label{epsdef}
\eps(t_f)= \frac{z_g -z_b}{z_g} 
 = 1- \frac{k~ t_f}{\sqrt{(1+A^2 k^2)}
   \,\,{\cal E}\left(k~ t_f,\sqrt{\frac{A^2 k^2}{1+ A^2 k^2}}\,\right) }\,.
\ee
We note that this $\eps$ and the one defined in eq.~(\ref{eps1})
are formally the same when the 
space-coordinate $x$ is replaced by the 
physical time-coordinate $t$.

The parameter $\eps$ depends very weakly on $t_f$ when many fluctuations are 
traversed, i.e. when $t_f \gg 2 \pi/k$.
In fact, we are free to choose $t_f$ according to eq.~(\ref{tint}).
With this choice,
the shortcut parameter depends only on the geometry $Ak$ 
of the brane fluctuation,
according to 
\be{}\label{epscomplete}
\eps=1- \frac{\pi/2}{\sqrt{(1+A^2 k^2)}\,\,
  {\cal E}^c\left(\frac{\pi}{2},\sqrt{\frac{A^2 k^2}{1+ A^2 k^2}}\,\right) }\,.
\ee
In the latter equation,
we have used the relation 
${\cal E}\left(n\pi,q\right)
= 2n\,{\cal E}^c\left(\frac{\pi}{2},q\right)$;
the expression 
${\cal E}^c\left(\frac{\pi}{2},q\right)$ 
is called the complete elliptic integral.
In the model developed here, 
an inference of $\eps$  
offers a direct measurement of the brane-fluctuation shape-parameter $Ak$.

The dimensionless ratio $Ak$ is essentially the aspect ratio (height to width) 
of the fluctuation.  For a brane with high (low) tension, we expect a 
low (higher) value of $Ak$.
Our assumption is that the brane tension is large, so that there is no 
curvature on large scales.  Accordingly, we expect a small value for 
$Ak$.  To first non-vanishing order in $Ak$, the parameter $\epsilon$ is
\be{}\label{orderAk}
\eps=\left(\frac{Ak}{2}\right)^2\,.
\ee
This approximation is valid until $Ak$ becomes of order unity,
after which $z_g \gg z_b$ and $\eps$ itself approaches unity.

An alternative to the periodic metric described here arises 
in spacetimes
in which the speed of light along flat 4D sections varies with the extra 
dimension coordinate $u$ due to different warp factors for the space and the 
time coordinates (``asymmetrically warped'' spacetimes)
\cite{freese,csaki}.
Such scenarios are realized, e.g. by a black-hole background in the bulk,
and may provide interesting consequences for the 
adjustment of the cosmological constant
\cite{csaki}.
A specific example 
is 
the two-brane scenario discussed in \cite{freese},
in which the 4+1 dimensional metric is given by
\be{}
ds^2 = dt^2 -[e^{-2 k u}a^2(t)\,d{\bf h}^2 + du^2].
\label{altmetr}
\ee 
Here ${\bf h}$ denotes an Euclidian three-vector. In this model a sterile 
neutrino could scatter out of our brane at $u_1$ on a geodesic perpendicular
to $h$, reach a second brane at $u_2>u_1$, and finally 
scatter off some fields confined on this second brane 
to return back to our brane.
It has been shown in \cite{freese} that again the on-brane
distance traveled in a given time interval
via a path on the hidden-sector brane and in the bulk can be larger than
the distance for pure on-brane travel in the same time interval.
We 
anticipate that the qualitative features of this and similar scenarios 
can be approximated
by the simple toy model discussed above.

\section{The two-state sterile-active oscillation probability}
\label{sec:twobytwo}
Let us illustrate the brane-bulk resonance for a 
simple system of one 
sterile neutrino $\nu_s$ and one active neutrino $\nu_a$.
The mass eigenstates are $m_2$ and $m_1$, respectively, in the sense
that for small mixing 
$\nu_s$ is mostly $\nu_2$ and $\nu_a$ is mostly $\nu_1$.

Ignoring the bulk for the moment, the evolution equation in flavor space reads
\be{}
i\frac{d}{dt}\left(\begin{array}{c} \nu_{a}(t)\\ 
\nu_{s}(t)  \end{array}\right) = 
H_F \left(\begin{array}{c} \nu_{a}(t)\\ 
\nu_{s}(t)  \end{array}\right),
\ee
and the Hamiltonian in the flavor basis is
\beq{}\label{Hflavor1}
H_F = E 
+ \half\,Tr 
+ \frac{\dmsq}{4\,E} \left(
\ba{rr}
-\cos 2\theta  & \sin 2\theta \\
\sin 2\theta & \cos 2\theta \\
\ea
\right)
\eeq
where $Tr=m_2^2 +m_1^2/2E$, $\delta m^2=m_2^2-m_1^2$ and 
$\theta$ is the mixing-angle in the unitary matrix relating 
flavor and mass bases:
\beq{}\label{U1}
|\nu_\alpha\rangle = U^*_{\alpha j}\,|\nu_j \rangle\,,
   \quad {\rm or} \quad U_{\alpha j} = \langle\nu_\alpha | \nu_j\rangle \,,
\eeq
with 
\beq{}\label{U2}
U=\left(
\ba{rr}
\cos\theta  & \sin\theta \\
-\sin\theta & \cos\theta \\
\ea
\right).
\eeq
%
We will call $\theta$ and $\delta m^2$ the standard mixing angle,
and standard mass-squared difference, respectively.
They are the analogs of vacuum values in MSW \cite{msw}
physics.

Now we let the sterile neutrino propagate in the bulk as well as on the brane.
If the brane were rigid and flat in its embedding, then the sterile geodesic 
is just the same as the active geodesic on the brane.
However, if the brane is curved in its embedding, as discussed above,
then the sterile neutrino may have a different 
trajectory,
with a shorter geodesic than that of the active neutrino constrained to the 
brane.
We will formulate this as an effective potential contributing to the 
sterile-sterile term of the Hamiltonian in flavor space.
Note that this is analogous to the Wolfenstein potential for the 
active-active term due to forward elastic scattering in matter, albeit with
three important differences.
The first is that the effective potential here
is the same for neutrino and antineutrino, because
gravitationally-determined geodesics do not distinguish between
particle and antiparticle.
The second difference is a more pronounced 
energy dependence here, with the effective mass-squared difference 
varying as $E^2$, not as $E$.
The third difference is that there is no time or space dependence
in the Hamiltonian.
We note that characteristics of 
this brane-bulk luminal/superluminal system  
resemble certain scenarios with Lorentz invariance violation
\cite{kostlsnd,lorentz}.

Due to the shortcut in the bulk (see the schematic in Fig.~\rf{fig:branebulk}),
the sterile state will appear to cover more distance on the brane than the 
active neutrino does in the same time, 
or equivalently, the same distance but in a shorter time.  
The ratio of apparent times at common distance 
or apparent distances at common time 
for the sterile and active neutrinos is 
$\delta t/t\simeq \delta z/z \simeq \eps$.
For the toy model
introduced in section~\ref{warpmetric} the parameter $\eps$ is given by
eq.\ (\ref{epscomplete}), or by eq.\ (\ref{orderAk}) to lowest 
non-vanishing order in $Ak$.

Adding the new contribution to the sterile-sterile element of the $H_F$, 
action/time$=E\,\frac{\delta t}{t}$,
and then removing the irrelevant energy and trace terms, 
we arrive at the effective Hamiltonian
\beq{}\label{Hflavor2}
H_F=
+ \frac{\dmsq}{4\,E}\, \left(
\ba{rr}
-\cos 2\theta  & \sin 2\theta \\
\sin 2\theta  & \cos 2\theta \\
\ea
\right)
+E\,\frac{\eps}{2}\,\left(
\ba{rr}
 1 &  0 \\
 0 & -1 \\
\ea
\right).
\eeq
The bulk term may beat against the brane term to give resonant mixing, 
i.e., for some energy $\Eres$ even a small standard angle can become large
or even maximal in the brane-bulk model.
%
The resonance condition is that the two diagonal elements in $H_F$ be equal,
which implies
\beq{}\label{Eres}
\Eres = \sqrt{\frac{\dmsq\,\cos 2\theta}{2\,\eps}}\,.
\eeq
%

Since the value of $\eps$ is unknown, the resonance energy could have 
almost any value, {\it a priori}.
However, if $\eps\ll 1$, as we assume, 
then we have the result $\dmsq\ll E^2_{\rm res}$.
Still, there is much parameter space available for resonance.
Our aim is to accommodate the LSND result in a four-neutrino 
framework, and so 
we will restrict the resonance energy with this in mind. 
It is worth noting that according to~(\ref{Eres}), 
a determination of $E_{\rm res}$ fixes $\eps$,
if $\delta m^2$ and $\cos 2\theta$ can be independently determined.
One way to independently determine $\delta m^2$ and $\cos 2\theta$ 
is to observe the active-sterile oscillation parameters far below 
resonance, where the oscillations are described by the standard formulas.
Note that knowledge of $\eps$, when available, yields the shape-parameter 
$Ak$ of the brane fluctuation, according to eq.~(\ref{orderAk}).

The value of $E_{\rm res}$ naturally divides the energy domain 
into three regions.
Below the resonance, oscillation parameters reduce to their standard values
and give the familiar oscillation results.
At resonance, the mixing angle attains a maximum
(but the effect on the oscillation probability  
can be reduced by a compensating factor in the 
$\delta m^2$ term).
Above resonance, the oscillations are suppressed.
Our strategy to accommodate the LSND data in a four-neutrino 
framework will be to set the resonant energy well below the CDHS data
to suppress oscillations for this experiment, but at or above the LSND 
energies, so as to not suppress (or even, to enhance) the LSND signal.

\begin{figure}
\centering
\includegraphics[clip,scale=1.0]{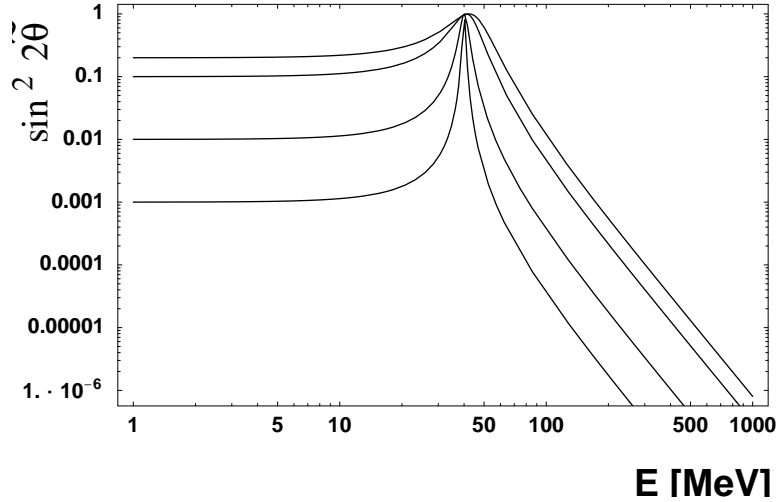}
\caption{Oscillation amplitude $\sin^2 2 \tilde{\theta}$
as a function of the neutrino energy $E_{\nu}$, for a resonance energy
of  $E_{\rm res}=40$~MeV. 
The different curves correspond to different values for the standard angle,
$\sin^2 2\theta=0.2, 0.1, 0.01, 0.001$ (from above).}
 \label{fig:ampemux}
\end{figure}

To find the new eigenvalue difference $\delta H$
and the new mixing angle $\twidtheta$ effected by the bulk,
one diagonalizes the $2\times 2$ system.
In terms of the new $\delta H$ and $\tilde{\theta}$ 
one obtains the usual expression for the flavor-oscillation 
probability
\be{}\label{oscprob}
P_{as}  =  \sin^2 2\twidtheta\;\sin^2 (\delta H\,D/2)\,,
\ee
with new values given in terms of standard values by 
\bea{}
\sin^2 2\twidtheta & = & 
\frac{\sin^2 2\theta}{\sin^2 2\theta +\cos^2 2\theta 
	\left[1-\left(\frac{E}{E_{\rm res}}\right)^2\right]^2}\,,
\label{osca}\\
\delta H &=&
\frac{\dmsq}{2\,E}\sqrt{ \sin^2 2\theta 
+\cos^2 2\theta \left[1-\left(\frac{E}{E_{\rm res}}\right)^2\right]^2}\,.
\label{osch}
\eea
The width of the resonance is easily derived from the classical 
amplitude in eq.~(\ref{osca}).\footnote{
A short calculation gives the Full Width in energy 
at a fraction $f$ of Maximum (FWfM) as
\beq{}\label{FWfM}
\frac{\Delta E({\rm FWfM})}{\Eres}=
  \left[1+\,\tan 2\theta\,\sqrt{\frac{1-f}{f}}\;\right]^{1/2}
- \left[1-\,\tan 2\theta\,\sqrt{\frac{1-f}{f}}\;\right]^{1/2}\,,
\eeq
which, for small $\theta$ reduces to 
$2\theta\,\sqrt{\frac{1-f}{f}}$.
Thus, the resonance is very narrow for a small standard angle.
For example, Full Width at Half Max is 
$\Delta E({\rm FWHM})= 2\,\theta\,\Eres$ for small angle. 
For a larger standard angle, the resonance becomes less dramatic.
}
Fig.~\ref{fig:ampemux} 
shows $\sin^2 2\tilde{\theta}$ for different values of $\sin^2 2\theta$
as a function of energy.

For $E\gg E_{\rm res}$, the sterile state decouples from the active state,
as $\sin^2 2\tilde{\theta} \rightarrow 0$.
Although the example presented in this section contains a single sterile state 
and a single active state, the decoupling of the sterile state(s) from the 
active state(s) is a general feature.

\section{Accommodating the LSND result}

\begin{table}
\begin{center}
\begin{tabular}{l|c|c|c|c}
& $\alpha \beta$ & $E_\nu$ & $D$ 
& $\sin^2 2 \tilde{\theta}_{\alpha \beta}$  \\
\hline
\hline
LSND      & $\mu e$  & 20-52.8~MeV  & 30~m & $>0.003$\\
KARMEN    & $\mu e$  & 20-52.8~MeV  & 17.7~m & $<0.002$\\
MiniBooNE & $\mu e$         & 0.1-1.0~GeV & 540~m & $\sim0.0006$\\
BUGEY     & ${e  \hspace*{-1mm}\not{\!e}}$  & 1-6~MeV & 25~m & $<0.15$\\
CDHS      & ${\mu\hspace*{-1mm}\not{\!\mu}}$ & $>1$~GeV & 755~m &  $<0.1$\\
\hline
\end{tabular}
\caption{Flavor channels, beam energies, 
oscillation distances (or differences of far and near 
detectors), and limiting oscillation amplitudes
in the large $\delta m^2$ limit, for the relevant experiments 
\protect{\cite{lsnd,lsnd-karmen,miniboone,bugey,cdhs}}. For BUGEY, the
25~m measurement has been chosen. 
\label{beamE}}
\end{center}
\end{table}

As the sterile neutrino mass is not protected by the gauge symmetry
of the Standard Model, it is natural to assume it to be larger than the
masses of the active neutrinos. We thus focus here on a $3+1$ neutrino 
spectrum \cite{3+1}, i.e. 
three active neutrinos are separated by the LSND mass-squared gap
$\delta m^2_{\rm LSND}$ from the dominantly sterile state
$\nu_4 \sim \nu_s$.

In the present model, $\delta \tilde{m}^2_{\rm LSND}$ differs from  
the standard formalism by the square-root factor in (\ref{osch}).
If the new resonance occurs in the energy range of LSND/KARMEN, 
this factor might allow a fit with a larger $\delta m^2$.
We do not pursue this subtlety in this work (see, however, the discussion
in section \ref{sec:wdm}).

For short distances, the mass-squared differences in the $3+1$ spectrum
can be taken as 
$\delta m^2 \equiv \delta m^2_{14} = \delta m^2_{24} = \delta m^2_{34}$,
with all other differences set to zero.
There results in all short-distance oscillation probabilities 
a universal factor of $\sin^2 (\delta H~D/2)$,
with $\delta H$ given previously in eq.~(\ref{osch}).
As a result of this universality, the relevant oscillation amplitudes
can be completely described using a two-neutrino formulation.
First, we define the linear combination of active flavors which couples to the 
heavy $|\nu_4\rangle$ mass eigenstate as $|\nu_a\rangle$.
Below the resonant energy, we write
\be{}\label{nu4}
|\nu_4\rangle=\cos\tilde{\theta}|\nu_s\rangle+\sin\tilde{\theta}|\nu_a\rangle\,.
\ee
Above the resonant energy, the mass eigenstates have to be relabeled
($4 \leftrightarrow 1$);
i.e. the isolated state contains little $|\nu_a\rangle$ but much 
$|\nu_s\rangle$.
Put another way, the oscillations considered still occur above the resonance
over the large active-sterile mass gap,
but effectively with the interchange $\cos \tilde{\theta} 
 \leftrightarrow \sin \tilde{\theta}$ in eq.~(\ref{nu4}).
We note that $\sin^2 2\tilde\theta$ is unaffected by this interchange.

To produce $\nu_\mu$-$\nu_e$ oscillations, it is necessary that 
this state $|\nu_a\rangle$ contains $|\nu_\mu\rangle$ and $|\nu_e\rangle$.  
It may also contain $|\nu_\tau\rangle$.
For simplicity, we will take this state to be a mixture of just 
$|\nu_\mu\rangle$ and $|\nu_e\rangle$:
\be{}
\label{nuactive}
|\nu_a\rangle = \cos\theta_* |\nu_\mu\rangle+\sin\theta_* |\nu_e\rangle\,.
\ee
Thus, we have 
\bea{}\label{Ue4}
\tilde{U}_{e4}&=&\sin\tilde{\theta} \,\sin\theta_*\,,\\
\label{Umu4}
\tilde{U}_{\mu 4}&=&\sin\tilde{\theta} \,\cos\theta_*\,,
\eea
below the resonant energy,
and $\cos \tilde{\theta} \leftrightarrow \sin \tilde{\theta}$
in eqs.~(\ref{Ue4}) and (\ref{Umu4}) above the resonant energy.

The oscillation amplitude relevant for the LSND \cite{lsnd},
KARMEN \cite{lsnd-karmen}
and MiniBooNE \cite{miniboone} appearance experiments
is given by
\be{}\label{stlsnd}
\sin^2 2\tilde{\theta}_{\rm LSND} = 
-4\sum_{j<4}
	\tilde{U}_{e4}\,\tilde{U}_{\mu 4}\,\tilde{U}_{ej}\,\tilde{U}_{\mu 4}
	= 4\,\tilde{U}_{e4}^2\,\tilde{U}_{\mu 4}^2
	= \sin^2 2\theta_*\,\sin^4 \tilde{\theta} \,.
\ee
Similarly, the oscillation amplitudes for the $\nu_e$ and $\nu_\mu$
disappearance experiments BUGEY \cite{bugey} and CDHS \cite{cdhs}
are given by
\bea{}
\label{BUGEYamp}
\sin^2 2 \tilde{\theta}_{e\not{e}}&=& 
4\,\tilde{U}_{e 4}^2 \,(1- \tilde{U}_{e 4}^2)
= 4\,\sin^2 \tilde{\theta} \sin^2 \theta_* \,
(1- \sin^2 \tilde{\theta} \sin^2 \theta_*),\\
\label{stcdhs}
\sin^2 2 \tilde{\theta}_{\mu\not{\mu}}&=& 4\,\tilde{U}_{\mu 4}^2 
\,(1- \tilde{U}_{\mu 4}^2)
= 4\,\sin^2 \tilde{\theta} \cos^2 \theta_* 
\,(1- \sin^2 \tilde{\theta} \cos^2 \theta_*),
\eea
respectively. 
The far-right expressions in these equations hold below the resonance;
above the resonance one must interchange 
$\cos \tilde{\theta} \leftrightarrow \sin \tilde{\theta}$,
as explained above.

Eq.~(\ref{BUGEYamp}) may be inverted to give 
$\tilde{U}_{e 4}^2 = \sin^2 \tilde{\theta}_{e\not{e}}$,
and eq.~(\ref{stcdhs}) to give
$\tilde{U}_{\mu 4}^2 = \sin^2 \tilde{\theta}_{\mu\not{\mu}}$.
Were these probabilities not energy-dependent, as in the standard 
vacuum case, one could substitute these results into
eq.~(\ref{stlsnd}) to get 
\be{}\label{smallangle}
\sin^2 2\theta_{\rm LSND}=
4\,\sin^2 \tilde{\theta}_{e\not{e}}\sin^2 \tilde{\theta}_{\mu\not{\mu}}
\simeq
\frac{1}{4}\sin^2 2\theta_{e \not{e}}\sin^2 2\theta_{\mu \not{\mu}}\,,
\ee
with the latter expression holding in the small angle approximation.
One recovers the well-known result in the standard case, 
that the LSND amplitude 
is doubly suppressed by stringent bounds on the
BUGEY and CDHS amplitudes. 
This fact excludes the standard 3+1 neutrino 
models from describing the results of all short-baseline neutrino
experiments \cite{schwetzfits}.
However, eq.~(\ref{smallangle}) is not valid in general in our bulk
shortcut scenario.
As we shall demonstrate, the energy-dependence imparted to the 
mixing-angles and to $\delta H$ by the bulk shortcut allows consistency.

It will also be useful to list the amplitudes for $\nu_\mu$-$\nu_s$ and 
$\nu_e$-$\nu_s$ oscillations, for these may affect atmospheric and 
solar oscillations.  
As shown in eqs.~(\ref{osca},\ref{osch}), the active-sterile 
oscillations are governed by vacuum values below the resonant energy, 
are maximal at the resonant energy,
and are suppressed above the resonant energy.  The same is true therefore for 
$\nu_\mu$-$\nu_s$ and $\nu_e$-$\nu_s$ oscillations:
\be{}\label{mu-s}
\sin^2 2 \tilde{\theta}_{\mu s} = \cos^2 \theta_* \,\sin^2 2 \tilde{\theta}\,,
\ee
and
\be{}\label{e-s}
\sin^2 2 \tilde{\theta}_{e s} = \sin^2 \theta_* \,\sin^2 2 \tilde{\theta}\,,
\ee
with  $\sin^2 2 \tilde{\theta}$ given in eq.~(\ref{osca}). 

For the analysis of neutrino oscillations in the bulk-shortcut scenario,
the energy of the neutrino beam is of crucial importance. 
In Table~\ref{beamE}, the relevant experiments are shown together
with the flavor amplitude to which they are sensitive, the neutrino beam 
energy, and 
the bound (or for LSND, the favored region) for the amplitude.  
Our task is to compare these experimental bounds 
(and for LSND, the positive signal) with the 
energy-dependent oscillation probabilities listed above.
Note that all experimental data must be accommodated with four 
parameters: the standard mass-squared difference $\delta m^2$,
the standard mixing angles $\theta$ (describing $\nu_a$-$\nu_s$ mixing) 
and $\theta_*$ (parametrizing the flavor composition of $\nu_a$),
and the shortcut parameter $\epsilon$ 
(or equivalently, the resonant energy $E_R$,
as given in eq.~(\ref{Eres})).

\subsection{BUGEY}
BUGEY detects reactor neutrinos of energies an order of magnitude 
below the energies of LSND.
We assume that the BUGEY energies are also far below the resonance energy, 
in which case the bulk shortcut effects decouple and the oscillation
amplitude is given by the standard limit of 
eq.~(\ref{BUGEYamp}),
\be{}
\sin^2 2 \theta_{e\not{e}}\simeq 4\,U_{e4}^2 
	\simeq 4\, \sin^2 \theta\, \sin^2 \theta_*.
\ee
Here the smallness of $\sin^2 2 \theta_{e\not{e}}$
has been assumed, in concordance with the BUGEY limit 
$\sin^2 2 \theta_{e\not{e}}<0.15$.
We note that the BUGEY amplitude maybe suppressed by a small 
$\sin^2\theta_*$, or small $\sin^2\theta$, or both.

In the following we discuss two parameterizations compatible with LSND
and CDHS which sufficiently suppress this BUGEY amplitude, 
namely 
$\sin^2 2 \theta=0.9$, $\sin^2\theta_*=0.01$, 
and
$\sin^2 2 \theta=0.45$, $\sin^2\theta_*=0.1$.
The first parameterization has a large active-sterile mixing, 
$\theta=36^\circ$, but still a small   
$\sin^2 2 \theta_{e\not{e}}=0.014$.
The second parameterization yields a moderate $\theta=21^\circ$ 
and small $\sin^2 2 \theta_{e\not{e}}=0.052$.
Since the BUGEY amplitude is sufficiently suppressed,
there is no bound from BUGEY data on the value of $\delta m^2$.

Future reactor experiments are proposed to search for nonzero $U_{e3}$.
In these experiments, mixing of the sterile state with $\nu_e$ would have 
a measurable effect, mimicking a non-zero $U_{e3}$. 
In contrast to the effects of a
non-zero $U_{e3}$ (equivalently,
a non-zero $\theta_{13}$) and as a result of the large oscillation
phase $\propto \delta m^2_{\rm LSND}$, 
the effect will be seen in both the near and the far detectors, though,
and can be as large as $\sin^2
2\theta_{13} = 0.05$.

\subsection{CDHS}
The accelerator oscillation experiment CDHS operated with
neutrino energies above a GeV. 
At energies $E\gg E_{\rm res}$ the
active-sterile mixing is suppressed, and one can approximate
for small $\sin \theta_*$ (from eqs.~(\ref{osca}) and (\ref{stcdhs}))
\be{}
\sin^2 2 \tilde{\theta}_{\mu\not{\mu}}
\simeq 
\cos^2\theta_*\,\sin^2 2 \tilde{\theta}
\simeq
\cos^2 \theta_* \tan^2 2 \theta 
\left(\frac{E}{E_{\rm res}}  \right)^{-4}\,.
\ee
This implies that neutrino oscillations in the CDHS experiment are suppressed
by a factor between $10^{6}$ and 40 for a resonance energy in the range
of 30-400~MeV, making the oscillation amplitude 
$\sin^2 2 \theta_{\mu \not{\mu}}$ unobservable above $\sim$~GeV,
even if the $\sin^2 2 \theta_{\mu \not{\mu}}$ were maximal below the resonance.

\subsection{LSND and KARMEN}
We have suppressed the BUGEY oscillation amplitude with the choice of a 
small $U_{e4}^2$ below resonance.
We have suppressed the CDHS oscillation amplitude with the choice of a 
resonant energy below 400~MeV.
This leaves two possibilities for the effect of the resonance on the 
LSND/KARMEN energy range, 20-53~MeV.  The resonance may occur within this 
range, in which case the effect is observable.
Alternatively, the resonance may occur above this range,
in which case there is little change from the standard prediction and fit.
We show below that either possibility can be realized with a resonance 
consistent with all other oscillation data. 
We also show below that the two different choices have very different 
consequences for the on-going MiniBooNE experiment.

From eq.~(\ref{stlsnd}) and the following discussion, 
we have for the LSND oscillation amplitude
\be{}
\sin^2 2 \tilde{\theta}_{\rm LSND}
=\frac{1}{4} \sin^2 2 \theta_* \,(1\mp \cos 2 \tilde{\theta})^2\,.
\label{amplsnd}
\ee
The sign of the $\cos 2 \tilde{\theta}$ term corresponds to energies
below and above $E_{\rm res}$, respectively,
accounting for the fact that the states have to be relabeled
when crossing the resonance so that
the oscillations considered occur over the large active-sterile mass gap.
This formula also applies for the KARMEN and MiniBooNE amplitudes.

\begin{figure}
\centering
\includegraphics[clip,scale=1.0]{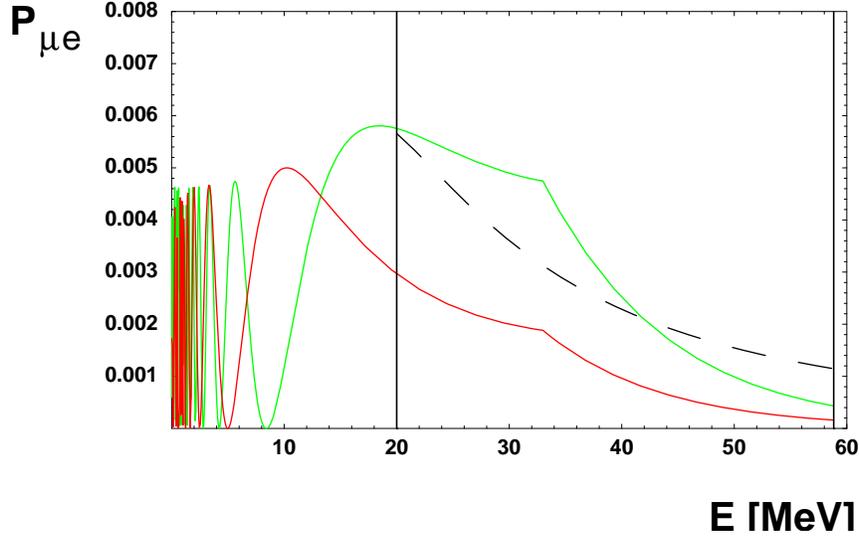}
\caption{Bulk shortcut oscillation probabilities for LSND (light/green) and 
KARMEN (dark/red) as a function of the neutrino energy.
Shown is a scenario with $E_{\rm res}=33$~MeV; $\sin^2 \theta_*=0.01$;
$\sin^2 2 \theta=0.9$; $\delta m^2=0.7$~eV$^2$.
For comparison, a standard oscillation probability for LSND 
($\delta m^2=0.8$~eV, $\sin^2 2 \theta_{\rm LSND}=0.006$) 
is displayed (dashed). 
The vertical lines indicate the energy window of LSND and
KARMEN. 
\label{eres33}
}
\end{figure}

\begin{figure}
\centering
\includegraphics[clip,scale=1.0]{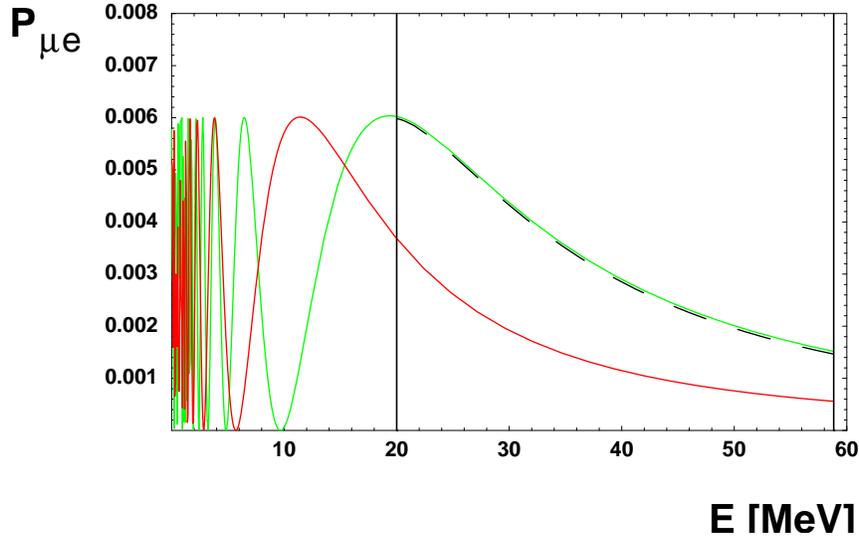}
\caption{
Bulk shortcut oscillation probabilities for LSND (light/green) and 
KARMEN (dark/red) as a function of the neutrino energy.
Shown is a scenario with $E_{\rm res}=400$~MeV,
$\sin^2 \theta_*=0.1$;
$\sin^2 2 \theta=0.45$; $\delta m^2=0.8$~eV$^2$.
For comparison, a standard oscillation probability 
($\delta m^2=0.8$~eV, $\sin^2 2 \theta_{\rm LSND}=0.006$) for LSND is 
displayed (dashed).
The vertical lines indicate the energy window of LSND and
KARMEN. 
\label{eres500}
}
\end{figure}

In Figs.~(\ref{eres33}) and (\ref{eres500}) 
the oscillation probability $P_{\mu e}$
predicted for LSND in the bulk-shortcut scenario 
is compared to the prediction of the standard oscillation case (dashed).
Also shown is the expectation for KARMEN. 
Two very different sets of parameters have been chosen: 
one scenario has a resonant energy $E_{\rm res}=33$~MeV
in the LSND/KARMEN energy range, 
and the other has a resonant energy $E_{\rm res}=400$~MeV
far beyond the LSND/KARMEN energy range. 

To maintain consistency between the LSND and KARMEN data,
it is necessary to exploit the differing distances of the two 
experimental configurations, $D_{\rm LSND}=30$~m and 
$D_{\rm KARMEN}=17.7$~m according to Table 1.
As in the standard approach, this is done as follows:
According to eq.~(\ref{oscprob}), the neutrino remains in its first 
oscillation until $\delta H\,D = 2\,\pi$.
For $\delta H\,D \ll 2\,\pi$, the factor $\sin^2 (\delta H\,D/2)$ 
is well approximated by just $(\delta H\,D/2)^2$, 
giving oscillation probabilities a quadratic dependence on distance.
Since the baseline for KARMEN is about half that of LSND,
choosing parameters such that 
$\delta H\,D_{\rm KARMEN}  \lsim 1$, 
with $\delta H$ given in eq.~(\ref{osch}),
suppresses KARMEN by a factor of four compared to LSND.
This allows a slice of LSND parameter space to remain viable,
in the face of the KARMEN null result.

The requirement for LSND/KARMEN neutrinos to remain within their 
first oscillation is the standard one,
$\delta m^2 \sim {\rm eV}^2$.
We adopt this value here.

At this point, we may invert eq.~(\ref{Eres}) to determine the value 
of $\eps$:
\be{}\label{epsis}
\eps=\frac{\cos 2\theta\,\delta m^2}{2\,E_{\rm res}^2}
	= \frac{\cos 2\theta}{2}\,\left(\frac{\delta m^2}{{\rm eV}^2}\right)\,
	  \left(\frac{100\,{\rm MeV}}{E_{\rm res}}\right)^2\times 10^{-16}\,.
\ee
The freedom for $E_{\rm res}$ in this model allows $\eps$ to range 
over $\sim 10^{-18}$ to $10^{-16}$.
According to eq.~(\ref{orderAk}), this in turn implies a 
shape-parameter (height to width ratio) for the brane fluctuation
of $Ak \sim 10^{-8}$. These parameter values have to be eventually
explained in a theory of brane dynamics.

Both choices for $E_{\rm res}$ exhibit a viable LSND energy spectrum,
as seen in the figures.
For the case where $E_{\rm res}\gg E_{\rm LSND}$,
the LSND/KARMEN analysis and fit is the standard one.
For the case where $E_{\rm res}$ lies in 
the LSND range 20~MeV$<E<53$~MeV,
the energy dependence of the oscillation amplitude is modified considerably.
For this latter case, we expect that the resonance should be 
evident in the LSND spectral data, and we encourage a re-analysis of the
measured LSND energy-spectrum by the collaboration.

\subsection{Solar and Atmospheric Data}
The active-sterile amplitudes for $\nu_\mu$ and $\nu_e$ are given 
in eqs.~(\ref{mu-s}) and (\ref{e-s}).
However, the same physics which suppresses active-sterile
oscillations in BUGEY data and CDHS data, also suppresses 
active-sterile oscillations in solar and atmospheric data,
respectively.
For solar oscillations,
the smallness of $|U_{e4}|^2$ evades the experimental constraint.
For atmospheric neutrinos,
the measured energies are above the resonant energy,
and so atmospheric oscillations into the sterile state are suppressed.
The event sample below 500~MeV may contain enhanced $\nu_s$ production,
but experimentally this would be difficult to confirm.

In fact, even the 2+2 model of four-neutrinos 
can be resurrected with the brane-bulk resonance.
In the same way that the suppression of sterile-active oscillations above 
the resonant energy neutralizes the CDHS constraint for the 3+1 model,
so does it neutralize the atmospheric constraint for the 2+2 model.

A detailed
calculation is required to determine the nearly unitary $3\times 3$ 
active-neutrino 
mixing-matrix that results when the sterile state decouples at high energy.
We do not pursue this here.
However, we expect that the freedom to partition the state $|\nu_a\rangle$ 
among the 
$|\nu_e\rangle$, $|\nu_\mu\rangle$, and $|\nu_\tau\rangle$ is sufficient to 
yield acceptable 
phenomenology.
For example, although we have taken $\langle \nu_\tau | \nu_a \rangle =0$ for 
simplicity,
a $\nu_\mu\leftrightarrow \nu_\tau$ interchange symmetry, known to be 
consistent 
with all present data, can be 
incorporated here by changing
$|\nu_\mu\rangle$ in eq.~(\ref{nuactive})  
to $|\nu_\mu '\rangle = \frac{1}{\sqrt{2}} (|\nu_\mu\rangle +
|\nu_\tau\rangle )$.

\subsection{MiniBooNE}
The requirement that $E_{\rm res}\lsim 400$~MeV, well below the CDHS energy 
of $\sim 1$~GeV, leaves the resonance energy in the 
MiniBooNE range, 0.1 to 1~GeV, or even below. We predict
that MiniBooNE 
should see no signal above $\sim 700$~MeV in this model.

\begin{figure}[!t]
\centering
\includegraphics[clip,scale=1.0]{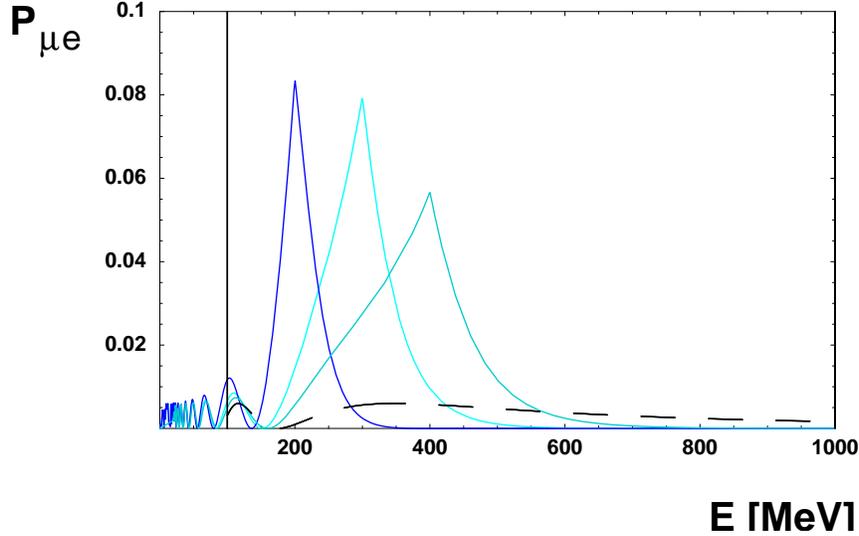}
\caption{
Bulk shortcut oscillation probabilities for MiniBooNE 
as a function of the neutrino energy.
Shown is a scenario with 
$\sin^2 \theta_*=0.1$;
$\sin^2 2 \theta=0.45$; $\delta m^2=0.8$~eV$^2$.
The resonance energy is varied, $E_{\rm res}=200,~300,~400$~MeV,
from left to right (light to dark).
For comparison, the expectation for a standard oscillation solution 
($\delta m^2=0.8$~eV, $\sin^2 2 \theta_{\rm LSND }=0.006$)
for LSND is displayed (dashed). 
The vertical line indicates the energy threshold of MiniBooNE.
 \label{minispek}}
\end{figure}

If  $E_{\rm res}$ falls in the MiniBooNE range above 100~MeV, 
then MiniBooNE should observe a strongly enhanced signal 
as evidence for the bulk-shortcut resonance.
Near the resonance region, the exact expression
(\ref{amplsnd}) applies for MiniBooNE. 
Results of this expression are shown in Fig.~\ref{minispek}
for resonance energies of 200, 300 and 400~MeV. 
As can be seen, the strongly enhanced 
oscillation probability is unmistakable.

On the other hand, 
if $E_{\rm res}$ lies below the MiniBooNE threshold energy,
then active-sterile mixing is strongly suppressed for MiniBooNE.
At energies $E\gg E_{\rm res}$, one uses eq.~(\ref{osca}) 
in eq.~(\ref{amplsnd}) to approximate
\be{}\label{MiniBooNE}
\sin^2 2\tilde{\theta}_{\rm MiniBooNE}\simeq \frac{1}{16} \sin^2 2 \theta_*
\tan^4 2 \theta \left(\frac{E}{E_{\rm res}}\right)^{-8}.
\ee
Thus a null result is predicted for 
MiniBooNE in the case of a resonance energy (as in our 33~MeV example)
below the MiniBooNE threshold of ${\cal O}(100)$~MeV.

However, if $E_{\rm res}$ is too low for an observable effect in 
MiniBooNE, a distortion in the LSND spectrum is expected 
(see Fig.~\ref{eres33}).
A strong $\nu_\mu$~disappearance signal also is predicted for an experiment 
using neutrinos from stopped pions, being proposed for the Spallation 
Neutron Source (SNS).
This we discuss next.

\subsection{Muon-neutrino disappearance at the SNS}
\label{subsec:SNS}

While the sterile neutrino
effectively decouples from the active sector at the 
CDHS energy and above, 
there is no suppression of the active-sterile mixing at and below the 
resonance.
Therefore, a significant effect is predicted for $\nu_\mu$ disappearance at 
lower energies.

Just such a lower energy $\nu_\mu$ disappearance experiment has been proposed~\cite{stoppedmu}
at the Spallation Neutron Source (SNS) being built at Oak Ridge.
The neutrino source would be  
stopped $\pi^+$'s, which undergo two-body decay to produce a  
monochromatic $\nu_\mu$ beam at 30~MeV.
In addition to the SNS source for stopped pions, 
there is the possibilty of a high-intensity ``proton driver'' at Fermilab
which would also include stopped pions on its physics agenda.
Detector distances at either site would be under 100~m from the pion source.
Thus, the $D/E$ is sufficiently small that only the LSND $\delta m^2$ can effect neutrino flavor change.

\begin{figure}[!t]
\centering
\includegraphics[clip,scale=1.0]{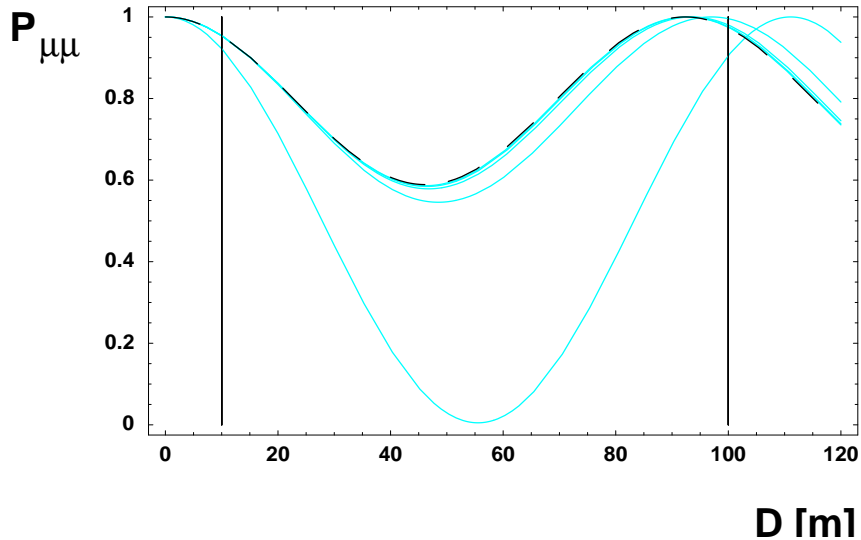}
\caption{
The muon-neutrino survival probability versus distance,
for a monochromatic neutrino-beam energy of 30~MeV from stopped pions.
The solid curves are parametrised by resonance energies 33, 100,
200 (identical for 300 and 400 MeV), in order of decreasing depletion.
The dashed curve is the result with no bulk shortcut.
The low-energy parameters for the 33 MeV resonance are chosen as in Fig.~3:
$\sin^2 \theta_*=0.01$; $\sin^2 2 \theta=0.9$; $\delta m^2=0.7$~eV$^2$;
while the parameters for the other curves are chosen as in Fig.~4:
$\sin^2 \theta_*=0.1$; $\sin^2 2 \theta=0.45$; $\delta m^2=0.8$~eV$^2$.
The vertical lines indicate the range of possible source to
near/far detector distances.}
 \label{fig:SNS}
\end{figure}

The amplitude for $\nu_\mu$-survival is given by eq.~(\ref{stcdhs}),
and the term oscillating with distance is 
$\sin^2 (\delta H\,D/2)$, with $\delta H$ given in eq.~(\ref{osch}).
The effects predicted for the stopped-pion  $\nu_\mu$ source at the SNS 
are shown in Fig.~(\ref{fig:SNS}).
The depletion of the $\nu_\mu$ beam due to substantial low-energy sterile-active mixing 
is considerable.
For $E\ll E_{\rm res}$, or for $\theta$ near maximal,
the oscillation length is insensitive to $E_{\rm res}$.
This explains the nearly common distance for the various minima in the figure.

We note that the large $\nu_\mu$-depletion in Fig.~(\ref{fig:SNS}) 
is specific to the parameters we have chosen,
and so should be interpreted as illustrative only.
Smaller mixing leads to smaller depletion.
If a $\nu_\tau$ component were added to the $\nu_a$ state, the 
$\nu_\mu$-depletion may be less.
Nevertheless, observable depletion of $\nu_\mu$'s from stopped pions is one 
of the more robust 
predictions of the brane-bulk model.

\section{Further implications for astrophysics and cosmology}

\subsection{BBN \label{bbnsec}}
Successful big-bang nucleosynthesis puts severe constraints on the 
equilibration between active neutrinos and even a single sterile state,
in the early Universe. The impact of active-sterile neutrino mixing
on nucleosynthesis is quite complex. 
It has been discussed extensively,
most recently in \cite{bbn}. 
A popular idea to resurrect LSND in view of the BBN constraints is to 
introduce a lepton asymmetry, 
which gives effective masses to the active neutrinos in
medium in the early universe, thus reducing the effective active-sterile
matter mixing angles. 
\footnote{Another way to reconcile BBN with the existence of a sterile neutrino
is to postulate a late-time phase transition \cite{Hannestad:2004px}.
With a reheating temperature of ${\cal O}(MeV)$ or less, the weak interaction 
simply 
does not have enough time to fully populate the neutrino modes.}

The bulk shortcut effect will further differentiate the sterile and 
active neutrinos.
It also may provide an alternative to a lepton asymmetry. 
Consider natural expectations for the evolution of the brane metric: 
A higher density in the early universe will lead to greater 
gravitational attraction, and so to more brane buckles;
and a larger temperature will increase thermal fluctuations of the brane.
At the epoch of BBN, the temperature is ten orders of magnitude larger
than today, and densities are thirty orders of magnitude larger than today.
In the alternative metric (\ref{altmetr}) mentioned earlier,
a higher density of scattering sites will increase 
the sterile neutrino scattering off from our brane. 
All these effects will increase the bulk shortcut parameter 
$\epsilon$ and thus reduce the resonance energy 
($E_{\rm res}\propto 1/\sqrt{\eps}$).
If $E_{\rm res}$ is reduced to a temperature near enough the BBN temperature of
$\sim 3$~MeV, earlier oscillations will be suppressed.
Our arguments that shortcuts had a larger $\epsilon$ and therefore smaller
$E_{\rm res}$ in the earlier Universe are consistent with Ishihara's statement,
that the magnitude of apparent causality violation on the brane increases with
increasing matter density \cite{ishihara}.

\subsection{Sterile neutrino mass and dark matter \label{sec:wdm}}
The present scenario offers an effective mechanism for sterile neutrino
production in the early Universe: 
when the temperature of the early universe drops below the
resonant energy, the active-sterile mixing becomes
maximal and active neutrinos are resonantly converted into sterile neutrinos.
Afterwords, active neutrinos are re-populated via reactions maintaining
thermodynamic equilibrium, until the active neutrinos decouple at energies
around 1~MeV.
The net effect, then, is to populate all neutrino modes,
sterile and active.

The occupation of states for a sterile neutrino with mass in the eV range 
impacts the effective total neutrino mass and number,
which in turn impacts the connection (``transfer function'')
between the cosmic microwave 
background anisotropy and today's large-scale structure.
A similar impact of neutrino mass/energy obtains for
measurements of galaxy bias stemming from galaxy-galaxy lensing, 
and for the large-scale power spectrum inferred from
Lyman-alpha forest observations in the Sloan Digital Sky Survey.
Specific consequences for cosmological evolution 
requires a detailed analysis (for recent works see \cite{Crotty:2003th}).
Here we just mention that, as described in the previous subsection 
\ref{bbnsec}, 
higher temperatures and densities in the early universe
may affect brane dynamics in a way to increase the effective shortcut
parameter and to reduce the resonant energy. This effect may keep the
sterile neutrino decoupled until the neutrino populations freeze out.

Since the effective 
$\delta \tilde{m}_{\rm LSND}^2\sim {\rm eV}^2$ is equal to
$\sim \delta m^2_{\rm LSND}\sin 2\theta$ on resonance, 
the true $\delta m^2_{\rm LSND}$
can be larger by $1/\sin 2\theta$.
In the case of a small mixing-angle, the gain can be very large.
With a sufficiently small $\theta$, 
the LSND sterile neutrino could play an important role
as warm dark matter (WDM) with a mass of order keV.
WDM has been proposed to solve the cuspy core problem
of cold dark matter scenarios \cite{wdm}.
Sterile WDM has also been proposed to 
induce the observed high-velocities of radio pulsars \cite{fuller}. 
Unfortunately, it seems difficult to fit the actual 
LSND energy spectrum with a small mixing-angle $\theta$,
as the required small mixing angles inducing sharp peaks at the 
resonance energy, 
see Fig.~\ref{fig:ampemux}.

\subsection{Supernova neutrinos}
Oscillations into sterile bulk neutrinos have interesting
consequences for supernova neutrinos.
For example, supernova cooling may be accelerated due to the emission
of Kaluza-Klein excitations of the sterile neutrino, resulting in constraints
on a product involving the sterile-active neutrino mixing and the 
radius of the extra dimension and/or a 
delayed explosion process \cite{Barbieri:2000mg}.
Since the efficiency of the bulk shortcut mechanism depends on the shortcut
parameter $\eps\simeq \left(\frac{Ak}{2}\right)^2$ with only $A$ being
bounded from the radius of the extra dimension, such constraints can always be
avoided by choosing a smaller $A$ and a larger $k$.

Furthermore, 
oscillations into sterile neutrinos
would affect r-process nucleosynthesis,
i.e.\ the rapid capture of neutrons on iron-sized seed nuclei, which is the
prime candidate for the synthesis of nuclei heavier than iron.
This process, which is believed to occur in type II supernovae,
is suppressed by $\nu_e$-capture on neutrons, which transforms
the target neutrons into protons, forming stable $\alpha$ particles. 
It has been shown that $\nu_e$-capture can be suppressed sufficiently 
if $\nu_e$'s oscillate strongly into sterile neutrinos
\cite{rproc}.

The bulk shortcut scenario will change this picture of 
r-process nucleosynthesis slightly.
First, resonances, now involving matter effects and bulk effects, 
may find their energies shifted toward smaller values by the bulk
shortcut. 
Moreover, 
supernova neutrinos with energies above the brane-bulk 
resonance will not
experience any level-crossing when propagating out 
of the supernova, resulting in a cutoff of the active-sterile neutrino
oscillation probability above $E_{\rm res}$.

\subsection{Horizon problem}
Finally,
the sterile neutrino could couple more strongly to brane fields than the 
graviton, especially in the resonance region around $E_{\rm res}$.
Thus, a solution to the horizon and homogeneity problems, proposed 
in \cite{kaelb,freese} but based on gravitons,
might turn out to be more effective if 
based on sterile neutrinos.
Moreover, while bounds on the size of extra dimensions from precision
measurements of the gravitational force law \cite{caldwell} impose stringent 
constraints on the gravitational horizon, these bounds may not be valid
for the extra dimensions felt by sterile neutrinos. The sterile neutrino
horizon thus may be even larger than the gravitational horizon.
Finally we stress that time dilation effects between the brane exit and the
brane re-entry points can lead to causality violations which increase the 
sterile neutrino horizon. Such effects will be discussed in a forthcoming 
paper \cite{pakv}.

\section{Discussion and Conclusions}
We have discussed active-sterile neutrino oscillation in an 
extra-dimensional brane world scenario. In such scenarios,
sterile neutrinos paths may take shortcuts in the bulk, 
which imparts an energy dependence to the oscillation amplitude. 
Resonant enhancement of active-sterile neutrino mixing
arises, parameterized by a shortcut parameter $\eps\equiv\delta t/t$.
If the resonant energy lies in the range 30~MeV to 400~MeV,
suitably chosen between the BUGEY and CDHS energies,
then all neutrino oscillation data can be accommodated in a consistent 3+1 
neutrino framework.
Such an energy range corresponds to $\eps$ in the range $10^{-18}-10^{-16}$,
and to brane fluctuations with a height to width ratio of $\sim 10^{-8}$.
The resonant energy might be identifiable in either the LSND spectral data
and the muon neutrino disappearance from a stopped-pion source, 
or in the soon-to-appear MiniBooNE data.

There are further interesting consequences for neutrino physics.
We have mentioned that
even the 2+2 model of active-sterile mixing 
can be resurrected with the brane-bulk resonance.
Finally, we have sketched only briefly
several interesting consequences for astrophysics and cosmology,
consequences which remain to be worked out in detail.

\section*{Acknowledgments}
We thank the CERN Theory Group (HP and TW) 
and the 
Centre for Fundamental Physics (CfFP) of the CCLRC
Rutherford Appleton Laboratory
(SP and TW) for kind
hospitality and support, and Bruce Berger,
Gautam Bhattacharyya, Marco Cirelli, Klaus Eitel, Steen Hansen,
John G. Learned, 
William C. Louis, 
Arthur Lue, Geoffrey Mills, Lothar Oberauer,
Sergio Palomares-Ruiz 
and Glenn Starkman for useful comments and discussions. 
This work was supported in part by US DOE under the grants DE-FG03-91ER40833
and DE-FG05-85ER40226.

\end{document}